\documentclass[aps,prb,showpacs,twocolumn,superscriptaddress]{revtex4-2}
\usepackage{amssymb}
\usepackage{graphicx}
\usepackage{appendix}
\usepackage{dcolumn}
\usepackage{bm}
\usepackage{amsmath}
\usepackage{xcolor}
\usepackage{float}
\usepackage{braket}
\usepackage[colorlinks,urlcolor=blue,anchorcolor=blue,linkcolor=blue,citecolor=blue,breaklinks=true]{hyperref}
\usepackage[T1]{fontenc}
\usepackage{array}
\usepackage{booktabs}

\begin{document}
\title{Second-order topological insulator in Bilayer borophene}

\author{Licheng Wang}
\thanks{These authors contributed equally to the work.}
\affiliation{National Laboratory of Solid State Microstructure, Department of Physics, Nanjing University, Nanjing 210093, China}
\affiliation{College of Physics Science and Technology, Yangzhou University, Yangzhou 225002, China}
\author{Ali Hamza Qureshi}
\thanks{These authors contributed equally to the work.}
\affiliation{College of Physics Science and Technology, Yangzhou University, Yangzhou 225002, China}
\author{Yi Sun}
\affiliation{College of Physics Science and Technology, Yangzhou University, Yangzhou 225002, China}
\author{Xiaokang Xu}
\affiliation{College of Physics Science and Technology, Yangzhou University, Yangzhou 225002, China}
\author{Xiaojing Yao}
\affiliation{College of Physics and Hebei Advanced Thin Films Laboratory, Hebei Normal University, Shijiazhuang 050024, China}
\author{Xinli Zhao}
\affiliation{College of Physics Science and Technology, Yangzhou University, Yangzhou 225002, China}
\author{Ai-Lei He}
\email{heailei@yzu.edu.cn}
\affiliation{College of Physics Science and Technology, Yangzhou University, Yangzhou 225002, China}
\author{Yuan Zhou}%
\email{zhouyuan@nju.edu.cn}
\affiliation{National Laboratory of Solid State Microstructure, Department of Physics, Nanjing University, Nanjing 210093, China}
\author{Xiuyun Zhang}
\email{xyzhang@yzu.edu.cn}
\affiliation{College of Physics Science and Technology, Yangzhou University, Yangzhou 225002, China}

\date{\today}

\newcommand*\mycommand[1]{\texttt{\emph{#1}}}
\newcommand{\red}[1]{\textcolor{red}{#1}}
\newcommand{\blue}[1]{\textcolor{blue}{#1}}
\newcommand{\green}[1]{\textcolor{green}{#1}}

\begin{abstract}
	As the novel topological states, the higher-order topological insulators have attracted great attentions in the past years. However, their realizations in realistic materials, in particular in two dimensional systems, remains the big challenge due to the lack of adequate candidates. Here, based on the first-principle calculation and tight-binding model simulations, we identify the currently \emph{existing}  bilayer $\alpha_{5}$-phase  borophenes as the two-dimensional second-order topological insulators, protected by the $C_{2}$-rotational symmetry. The formation of interlayer B-B covalent bonds, stabilizing the bilayer borophenes and opening the large direct bulk gaps ($\sim 0.55-0.62$ eV) at Fermi level, plays the key roles.  The second-order topology is characterized by the bulk quantized quadrupole momentum. Our results enriches the candidates for the second-order topological insulators, and also provide a way to study topological states in borophenes.
	
\end{abstract}

\maketitle


\section{Introduction}

The emergent topological states, beyond the traditional Landau paradigm, such as the quantum Hall states~\cite{qh1,qh2}, quantum spin Hall states~\cite{ref1,ref2,ref3}, quantum anomalous Hall states~\cite{ref4,ref5,ref6,ref7,ref8,ref9}, topological crystalline insulators ~\cite{ref11,ref12}, etc., have become a fairly active research field in condensed matter physics, and are proposed to be potential in future electronic and spintronic devices. The recent proposed high-order topological insulators (HOTIs)~\cite{HOTI1,HOTI2,HOTI3,HOTI4,HOTI5,HOTI6,HOTI7,HOTI8,HOTI9,HOTI10} further extend the concept of topological state of matter. Differing from the conventional (first-order) \textit{d}-dimensional topological insulators, which host insulating bulk state but gapless edge states propagating along their $(d-1)$-dimensional boundary, the HOTIs feature the unique bulk-boundary correspondence, that is, the gapless edge states appear on the (\textit{d}-\textit{n}) ($\textit{n}>1$)-dimensional boundary in a $n$th-order topological insulator, while for otherwise, the states are gapped. Such novel topological states has attracted great interesting from both theoretical and experimental researches.

Experimentally, the $3$-dimensional HOTIs, characterized by the topologically protected $1$-dimensional hinge states (second-order) or $0$-dimensional corner states (third-order), have been realized in a few electronic~\cite{ref15,ref16,ref17}, acoustic~\cite{ref18}, and optical~\cite{ref19} systems, etc. In principle, the $2$D second-order topological insulator (SOTI) can be achieved by introducing a SSH-like mechanism in the first-order topological insulators. Following this idea, a number of $2$D SOTIs have been confirmed in the theoretical models~\cite{ref20,ref21,ref22,ref23,ref24}, and several candidates in realistic materials have also been proposed~\cite{ref25,ref26,ref27,ref28,ref29,ref30,ref31,ref32,ref33,ref34,ref35,ref36}. The first theoretical candidate for 2D SOTI is the graphdiyne predicted by Sheng \emph{et al.}, they found that the topologically protected $0$D corner states are robust against the symmetry breaking {perturbation and boundary geometric details~\cite{ref25}. By calculating the quantized fractional charge, Qian \emph{et al.} revealed that a series of $2H$ group-VIB transition metal dichalcogenides (TMD) $MX_{2}$ (M$=$Mo, W; X$=$S, Se, Te) are large-gap $2$D SOTIs protected by $C_{3}$ rotational symmetry~\cite{ref34}.  Recently, Mao \emph{et al.} proposed the electric control of higher-order topology in In$_{2}$Se$_{3}$ by the coupling of nontrivial corner states and $2$D ferroelectrics~\cite{ref36}. However, compared with the theoretical progresses, no experimental discovery of $2$D SOTIs have been reported yet. Moreover, the variety of theoretically designed $2$D HOTI candidates is still scarce, some of them are partially property-limited due to the fact that topological corner states are far from the Fermi level. Therefore, searching for the $2$D SOTI in realistic materials and understanding the formation mechanism remain a great challenge.

As an analogue of graphene or other carbon-based $2$D isomers, which display rich topological properties~\cite{ref25,ref27}, $2$D borophenes exhibit a diversity of geometries properties with various configurations of vacancies~\cite{ref37,ref38}. These borophenes exhibit an abundance of physical and chemical properties. However, the topological properties in borophenes have not received enough attentions. Previous studies have shown that graphene with porous structures can host the nontrivial high-order topology~\cite{ref39,ref40}. Then, a question rises whether the novel second-order topology can be realized in porous borophenes.

In this work, we identify the previous experimentally synthesized stable bilayer borophenes composed of $\alpha_{5}$-phase borophene (named $\alpha_{5}$-BLB) as the $2$D SOTIs, protected by the $C_{2}$ rotational symmetry. These stacked configurations are found to be thermodynamically stable with positive phonon frequencies and large negative formation energies due to the formation of interlayer B-B covalent. Differing from the metallic $\alpha_{5}$-phase monolayer, all the $\alpha_{5}$-BLB are semiconductors hosting sizable bandgaps (larger than $0.5$ eV) near the Fermi level. The second-order topological nature are characterized by the nonzero quantized quadrupole momentum in bulk and the robust corner states in nanodisks. We further show that the corner states are robust against the edge defects and weak $C_{2}$-breaking perturbations.

\section{Computational methods and model}

All the first-principle calculations are performed with projector-augmented wave (PAW) methods within Vienna ab initio Simulation Package (VASP)~\cite{ref43,ref44}. The exchange-correlation potential is adopted by the generalized gradient approximation (GGA) of Perdew-Burke-Ernzerhof (PBE) functional~\cite{ref45}. The cutoff energy is set as $450$ eV. The energy and force convergence criteria were set to be $10^{-6}$ eV and $0.01$ eV/\AA, respectively. In our calculation, a vacuum slab as large as $20$ \AA \ is applied to avoid the physical interactions of adjacent cells. For geometry optimization, the first Brillouin zone is sampled by $\Gamma$-centered $k$ mesh of $15\times 15\times 1$. A denser $k$-point grid of $25\times 25\times 1$ is used for electronic structure calculations. Phonon dispersions are calculated using $3\times 3\times 1$ supercells by Phonopy code on the basis of DFPT~\cite{ref46}. The first-principles Wannier Hamiltonian are constructed in Wannier$90$ package with the maximally localized Wannier functions~\cite{ref47,ref48}. The edge states are calculated by the WannierTools package \cite{ref49}, and the parities are calculated by IRVSP package~\cite{ref50}.


 We further construct the tight-binding Hamiltonian fitted from the first-principle calculations to explore its physical properties and the corresponding Hamiltonian is,
\begin{equation} \label{E1}
	H=\sum_{(i,j),\ell} t_{ij} c^{\dagger}_{i\ell}c_{j\ell}+\sum_{(i,j),\ell^{\prime}\neq\ell} t^{\prime}_{ij}c^{\dagger}_{i\ell}c_{j\ell^{\prime}}+\sum_{i,\ell} V_{i}c^{\dagger}_{i\ell}c_{i\ell}.
\end{equation}
Here, $c^{\dag}_{i\ell}$ is the electron creation operator at site $i$ with layer index $\ell$. $t_{ij}$, $t^{\prime}_{ij}$ and $V_{i}$ represent the intralayer hopping, interlayer hopping and on-site potential, respectively.
\section{Results and discussion}

\subsection{Monolayer $\alpha_{5}$-phase borophene}
We start from the flat $\alpha_{5}$-borophene monolayer. Its geometry is shown in Fig~\ref{Fig1}(a), exhibiting a triangular lattice composed of eleven B atoms per unit cell with the space group $Cmmm$. After optimization (Fig. S1 in Supplementary Materials (SM)), the structure is no longer flat with a small thickness fluctuation about $0.4$ \AA. Here, we mainly focus on the flat $\alpha_{5}$-borophene monolayer, which preserves the inversion symmetry $\mathcal{P}$ and time-reversal symmetry $\mathcal{T}$. For simplicity, we only show the B-\emph{p} components since they dominate the near-Fermi level bands. There are $11\times 3$ bands due to three \emph{p} orbitals in each B atom and eleven B atoms in each unit cell. The electronic band structure of flat $\alpha_{5}$-borophene monolayer (Fig~\ref{Fig1}(b)) reveals that the system is metallic with the energy bands crossing the Fermi level, similar to many monolayer borophenes~\cite{b1,b2}. Moreover, a Dirac-like crossing point appears at the high-symmetry $K$ point near the Fermi level, resembling the semimetal state in graphene and other materials~\cite{grapheneband,dirac1,dirac2}.

\begin{figure}[!htb]
    \begin{center}
        \includegraphics[width=0.9\linewidth]{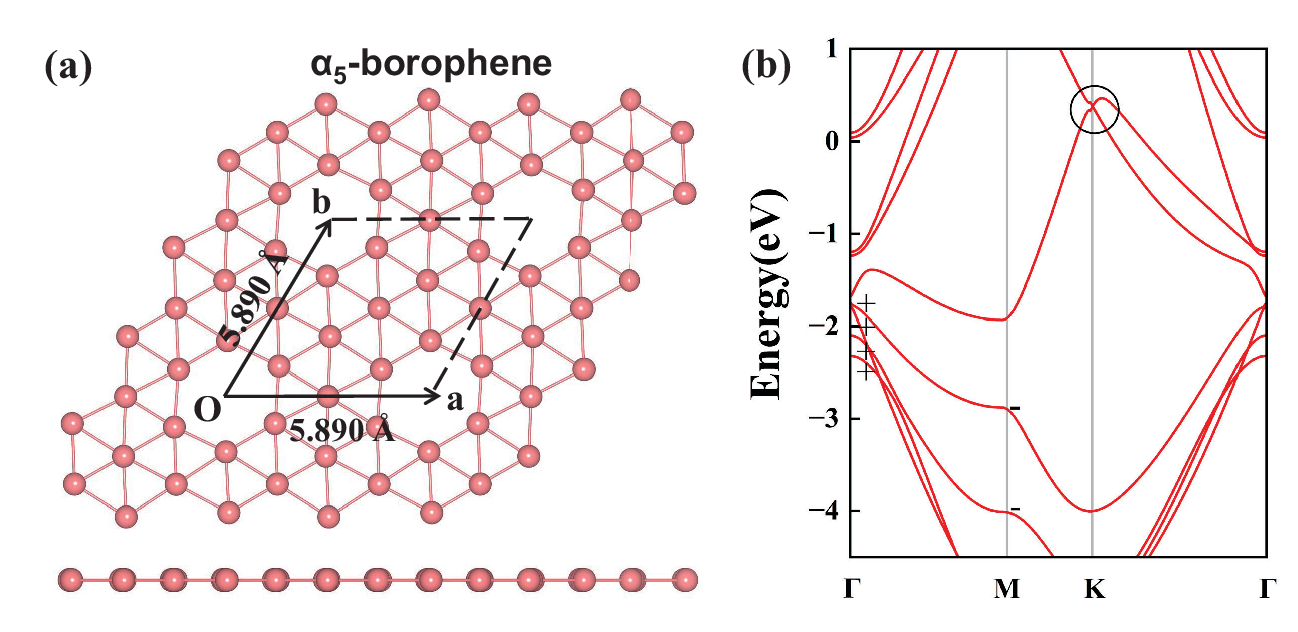}
    \end{center}
	\caption{(a) Top and side view of crystal structure of flat $\alpha_{5}$-phase boron monolayer. The black arrows represent the primitive lattice vectors. (b) The near-Fermi-level electronic band structure of $\alpha_{5}$-borophene. Part of parity values at $\Gamma$ and $M$ are marked. The black circle highlight the existence of Dirac-like crossing point in bulk.}
\label{Fig1}
\end{figure}

Interestingly, an indirect band gap opens in the $15$th and $16$th bands along the $k$-path (including only \emph{p}-orbitals ), indicating the existence of possible semiconductor phase in the system by heavy carrier doping. We explore whether there exists nontrivial band topology in the monolayer $\alpha_{5}$-borophene through the symmetry properties of $15$ occupied bands. The  $M_{1}$ and $M_{2}$ points are equivalent, but $M_{3}$ is not. Therefore, we calculate the eigenvalues of inversion operator $\mathcal{P}$ at high-symmetry $\Gamma$ and three $M$ points~\cite{ref62}, and find $n_{+}^{\Gamma} = 9/ n_{-}^{\Gamma} = 6$ at $\Gamma$ point and  $n_{+}^{M} = 7/n_{-}^{M}= 8$ at $M_{3}$ point, indicating a nontrivial topology with double band inversion. Unfortunately, the flat monolayer $\alpha_{5}$-borophene is proved to be unstable, which largely limit its potential application. The imaginary frequency in the phonon spectrum reveal that the optimized monolayer $\alpha_{5}$-phase boron monolayer (Fig. S1) is dynamically unstable.

\subsection{Bilayer $\alpha_{5}$ borophene}
Alternately, the stacking of multiple layers is a common strategy to stabilize the monolayer structure~\cite{ref51,ref52,ref53}. Moreover, the multilayer structure has the potential to open a global gap at the Fermi level, for example in bilayer graphene with external electric field~\cite{Rev1,Rev2}. We therefore construct the $\alpha_{5}$-BLBs containing $22$ B atoms in each unit cell. To determine the most stable stacking configuration, three types of stacking configurations for $\alpha_{5}$-BLB, namely the AA-stacking, AB-stacking, and AC-stacking, are constructed as shown in Fig.~\ref{Fig2}. Our results show that the space groups of $\alpha_{5}$-BLB with AA-, AB-, and AC-stacking are $Cmmm$, $C2/m$ and $C2$, and the optimized lattice constants are $5.709$ \AA, $5.708$ \AA \ and $5.70$ \AA, respectively. All three $\alpha_{5}$-BLBs have the time-reversal symmetry, and for AA-stacking, the $C_{2z}$ symmetry is preserved as well, which makes the present $\alpha_{5}$-BLB different from other $C_{n}$-symmetric SOTI candidates proposed before~\cite{ref25,ref27,ref34,ref35,c4symm}. The stabilizing of these $\alpha_{5}$-BLBs is accompanied with the strong interlayer interaction, that is, the strong interlayer boron-boron covalent bonds are formed (lower panels in Fig.~\ref{Fig2}), which distinguishes them from those traditional van der Waals (vdW) bilayers with weak interaction~\cite{ref54,ref55,ref56}. As a result of vacancies and interlayer bonds, the geometric symmetry of $\alpha_{5}$-phase borophene is broken and the intralayer B-B bonds are no longer uniform. The interlayer distances $d$ of the ``vdW'' interacting area of the $\alpha_{5}$- BLBs are around $3.31-3.40$ \AA, while those in the ``bonded'' area are much shorter, around $1.72-1.75$ \AA, which evidences the interlayer bonding characters in the bilayers. We further show the charge density difference of $\alpha_{5}$-BLB relative to the monolayer case in three stacking configurations, where the interlayer B-B bonds accumulate largest charges, indicating that they indeed form strong covalent bonds.

\begin{figure}[!htb]
    \begin{center}
	\includegraphics[width=0.9\linewidth]{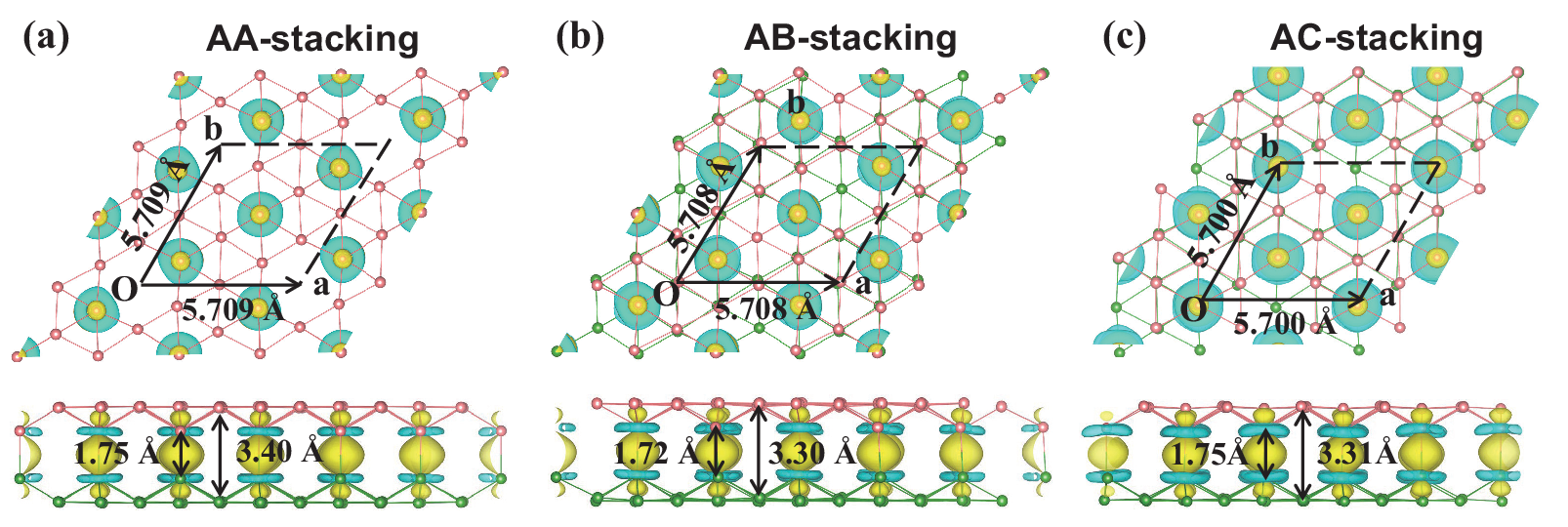}
    \end{center}
	\caption{Top and side view of crystal structures. Upper panels: (a)-(c) AA-, AB- and AC- stacking $\alpha_{5}$-BLB. The pink and green spheres represent B atoms of upper and lower layers, respectively.  Lower panels: The charge density difference relative to the monolayer case. The yellow and blue color indicate the gain and loss of charge. The long and short black arrows show the length of B-B bonds in the ``vdW'' area and ``bonded'' area, respectively. }
 \label{Fig2}
\end{figure}

To clarify the stability of the $\alpha_{5}$-BLB, we calculated their binding energies ($E_{b}$) and formation energies ($E_{f}$) for three configurations using following equations,
\begin{eqnarray}
	E_{b} &=& (E_{bilayer}-2 \times E_{monolayer})/N_{1}, \nonumber\\
	E_f &=& (E - n \times \mu_{B})/N_{2} .
\end{eqnarray}
Here, $E_{bilayer}$ and $E_{monolayer}$ are the energies of bilayer and monolayer $\alpha_{5}$-borophene, $\mu_{B}$ is chemical potential of B atom. $N_{1}$ represents the atom number of $\alpha_{5}$-BLB per unit cell, and $N_{2}$ is the atom number of $\alpha_{5}$ borophene monolayer (bilayer) per unit cell. The resultant binding energies for AA-, AB-, and AC- stacking configuration are $-0.27$ eV, $-0.22$ eV, and $-0.25$ eV per atom, respectively, indicating that the strong interlayer interaction of these BLBs. The calculated formation energies for three stacking configurations are $-0.67$ eV, $-0.65$ eV, and $-0.65$ eV per atom for AA-, AB-, and AC- stacking configuration, respectively, significantly larger than that of $\alpha_{5}$-phase monolayer borophene ($-0.381$ eV). Furthermore, no imaginary frequency along the high symmetry directions is observed in the phonon spectrums of all $\alpha_{5}$-BLBs (Fig. S2), in sharp contrast with that in the monolayer counterpart (Fig.S1), indicating their dynamical stabilities. Therefore, we conclude that the interlayer B-B bonds plays a crucial role on the dynamical stability and the electronic properties of the structures. Here, we would like to point out that the $\alpha$-phase of bilayer borophene had been recently synthesized in experiment~\cite{Bsynthesis}.

\begin{figure}[!htb]
	\begin{center}
		\includegraphics[width=\linewidth]{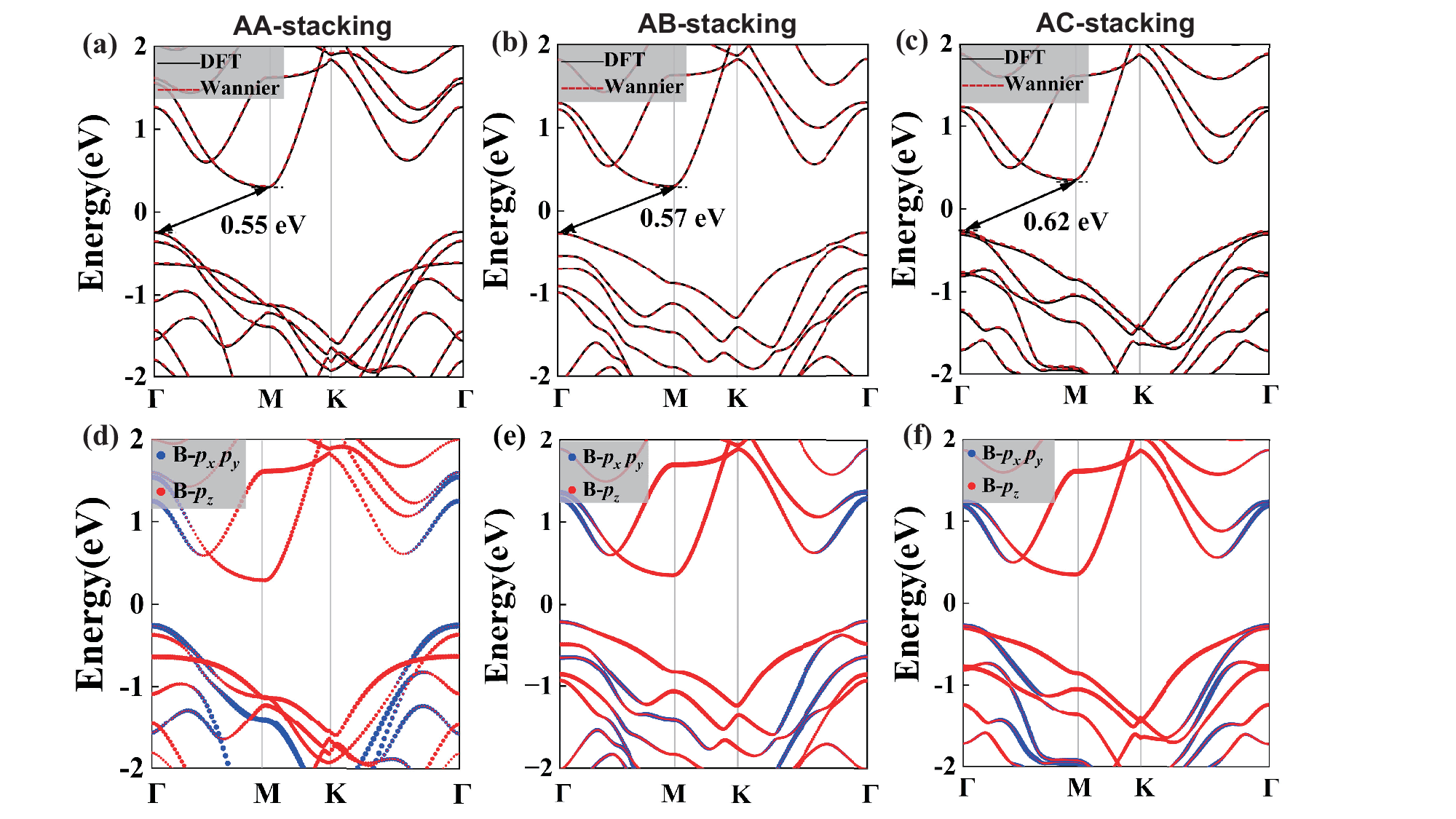}
	\end{center}
	\caption{Upper panels are the band structures of (a) AA-, (b) AB- and (c) AC-stacking $\alpha_{5}$-BLB along high symmetric paths obtained from DFT and Wannier functions (only B-$p_{x}$, $p_{y}$, and $p_{z}$ orbits are shown), respectively. Solid black and red dashes indicate DFT and Wannier functions. Lower panels are the corresponding orbital-resolved band structures. The blue and red indicate the $p_{x}+p_{y}$, and $p_{z}$ states of boron, respectively. The Fermi levels are fixed at $0$. }
	\label{Fig3}
\end{figure}

\subsection{Bulk bands and edge states}
The energy band structures of three stacked $\alpha_{5}$-BLBs are plotted in Fig~\ref{Fig3}(a)-(c). The spin-orbit coupling effect of $\alpha_{5}$-BLB is ignored due to its weakness. Clearly, three stacked $\alpha_{5}$-BLBs are all insulators with global band gaps $0.55-0.62$ eV near the Fermi level, well agreeing with the previous $\alpha$-phase BLB on Ag(111)~\cite{ref60}, while in sharp contrast with the metallic feature in $\alpha_{5}$-phase monolayer and other borophenes~\cite{ref57,ref58,ref59}. The orbital resolved bands near the Fermi level are shown in Fig~\ref{Fig3}(d)-(f), where the bands near Fermi level are dominated by the B-\textit{p} states. These results are obviously different from that of graphene or graphyne~\cite{grapheneband,graphyne}, in which the low energy bands are mainly contributed by the \textit{$p_{z}$} electrons of C atom. We further fit the tight-binding parameters with the maximally localized Wannier functions from the first-principle calculations for three $\alpha_{5}$-BLBs. We consider three \emph{p} orbitals per B atom in our models, yielding a $66$-band tight-binding model (Eq.~\ref{E1}). The fitted band structures are in good agreement with those from our first-principle calculations. To explore the origin of the large band gaps in these $\alpha_{5}$-BLBs, we examined the evolution of the band gaps as a function of interlayer spacing $d$ (see Fig. S3). The band gap of three $\alpha_{5}$-BLBs decrease gradually with increasing layer spacing, and completely closes when the interlayer distance increase to $4.01$ \AA, $3.94$ \AA \ and $4.18$ \AA, respectively. Therefore, We conclude that the band gaps of these $\alpha_{5}$-BLBs inherit from the formation of interlayer B-B bonds, in agreement with the previous reports~\cite{ref60}. Since the $\alpha_{5}$-BLBs in three stacking configurations display similar electronic energy bands, we focus on the results of AA-stacking $\alpha_{5}$-BLB for simplicity.

Further, the edge states of AA-stacking $\alpha_{5}$-BLB are examined with specified open boundaries. Here, four types of zigzag edges (ZZ-\textrm{I}, ZZ-\textrm{II}, ZZ-\textrm{III} and ZZ-\textrm{IV}) and one nearly flat edge are taken into consideration, and the corresponding edge states are illustrated in Fig~\ref{Fig4} (ZZ-\textrm{I}) and Fig.~S4 (ZZ-\textrm{II, III, IV}). The absence of gapless edge states indicate that the system is not a first-order topological insulator. However, the emergence of isolated edge states in the bulk energy gap suggests the potential existence of higher-order topological insulating states~\cite{ref25}.
\begin{figure}[!htb]
    \begin{center}
	\includegraphics[width=\linewidth]{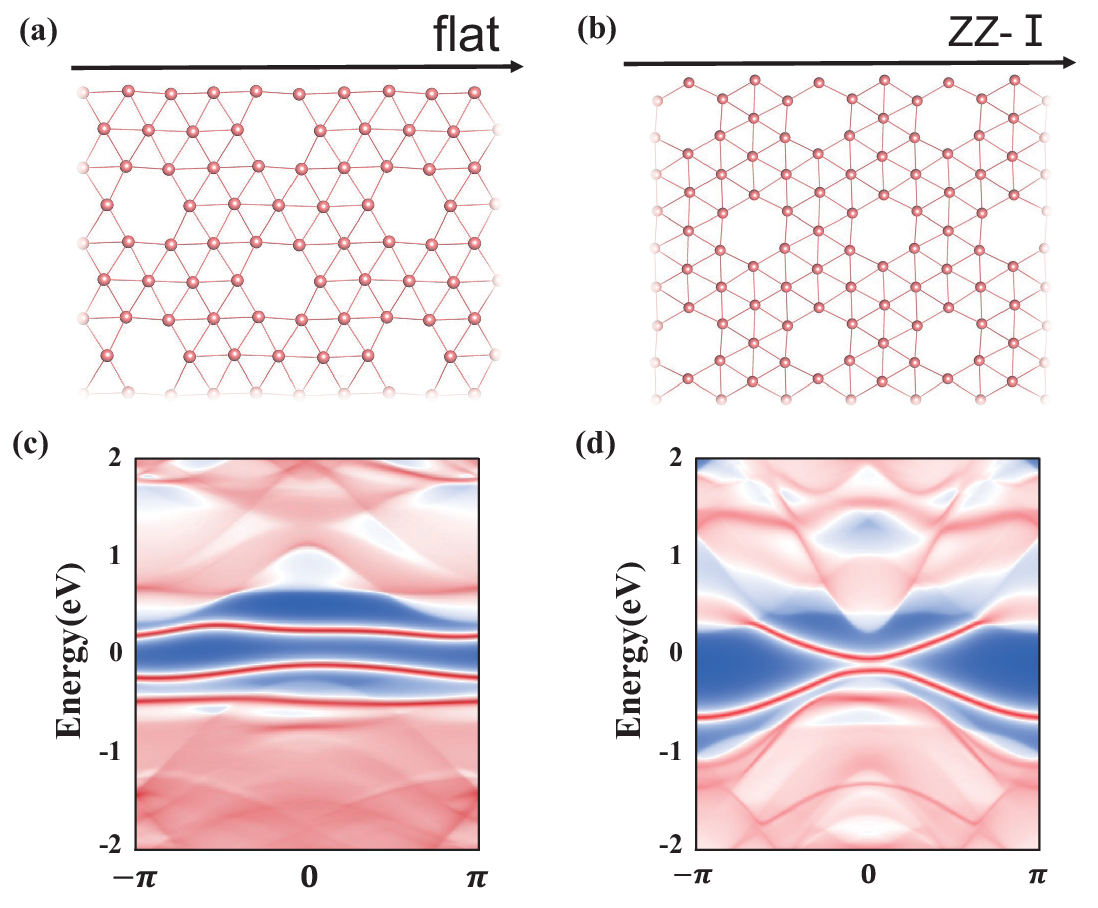}
\end{center}
	\caption{(a) and (b) Top views of the structure of AA-stacking $\alpha_{5}$-BLB. The black lines mark a near flat edges and a zigzag edge. (c) and (d) The edge states of the cylinder geometry with ZZ-\text{\textrm{I}} edge and flat edges, calculated by the Wannier interpolation method. }
\label{Fig4}
\end{figure}

The AA-stacking BLB is a spinless system preserving the time-reversal symmetry $\mathcal{T}$ and inversion symmetry $\mathcal{P}$. This naturally yields the vanishing of Chern number $\mathcal{C}=\frac{1}{2\pi}\int_{\text{BZ}} d\mathbf{k}^{2}\mathcal{F}(\mathbf{k})$ since the Berry curvature satisfies both $\mathcal{F}(-\mathbf{k})=-\mathcal{F}(\mathbf{k})$ and $\mathcal{F}(-\mathbf{k})=\mathcal{F}(\mathbf{k})$~\cite{ref61}. To further determine the higher-order topological properties of the BLB system, we calculate the dipole invariant ($\mathbf{p}=(p_{x},p_{y})$) and quadrupole momentum ($Q$) with following formulas~\cite{ref63}:
\begin{eqnarray}
	&& p_{i} =\frac{1}{2}\left(\sum_{n}2 p_{i}^{n} \ \text{mod} \ 2\right), \quad p_{i}^{n}=\frac{1}{2}q^{n}_{i}, \nonumber\\
   && (-1)^{q_{i}^{n}} =\frac{\eta^{n}(M)}{\eta^{n}(\Gamma)}.
\end{eqnarray}
Here, $i = x$ or $y$ indicates the reciprocal lattice vector, $\eta^{n}(M)$ and $\eta^{n}(\Gamma)$ are the parities of the $n$th occupied band at $M_{3}$ and $\Gamma$ point. The quadrupole momentum \textit{Q} is defined by the dipole momentum $p_{i}^{n}$~\cite{Quadrupole},
\begin{equation}
	Q_{ij} = \frac{1}{2}(\sum2p_{i}^{n}p_{j}^{n}\ \text{mod} \ 2).
\end{equation}
Our results show that the dipole momentum of the AA-stacking $\alpha_{5}$-BLB is $\mathbf{p}=(0, 0)$, and the quadrupole momentum is quantized with $Q_{ij}$ = 1/2, suggesting the emergence of SOTI state rather than the first-order topological insulating state in the AA-stacking $\alpha_{5}$-BLB.

\begin{figure}[!htb]
    \begin{center}
    \includegraphics[width=\linewidth]{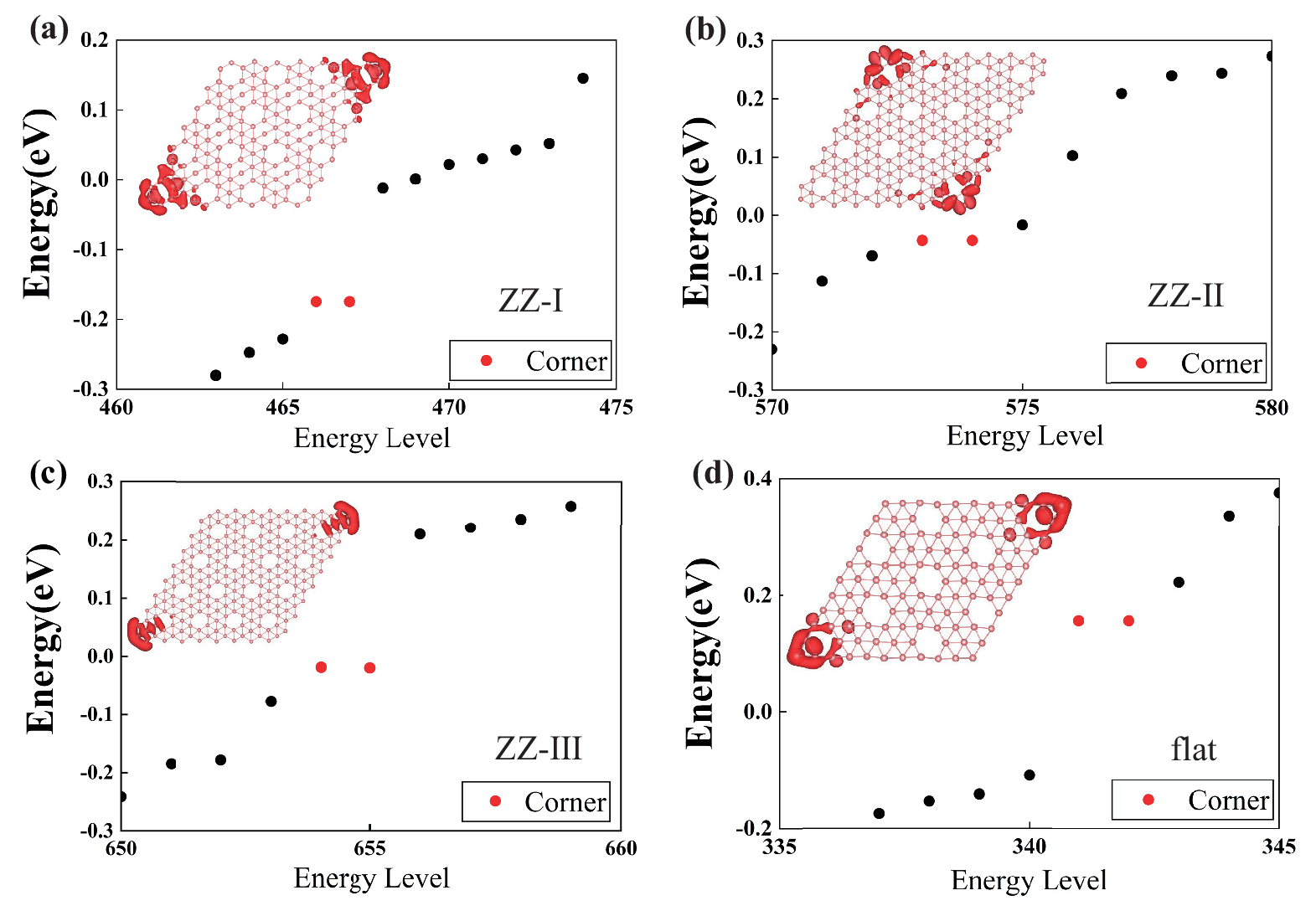}
    \end{center}
	\caption{The energy spectrum of four nanodisks with (a), (b) and (c) zigzag and (d) flat edges of AA-stacking $\alpha_{5}$-BLB calculated by DFT. The corner states are marked by red dots. The corresponding insets show the real-space distribution of the corner states. }
 \label{Fig5}
\end{figure}

\subsection{Robust corner states}
To further support the second-order topology of AA-stacking $\alpha_{5}$-BLB, we calculate the energy spectrums for several quadrilateral nanodisks with four edge configurations to search for the $0$D corner states. In all constructed edge configurations, there always exist two robustly degenerate states (marked as red dots) near the Fermi level within the gap of bulk and edge states (Fig.~\ref{Fig5}). Remarkably, the real-space distribution of the two degenerate states is well located at the two corners of the nanodisks, evidencing the SOTI feature in the $\alpha_{5}$-BLB. Due to the absence of particle-hole symmetry in real materials and the size effect of nanodisks, the corner states are not strictly at zero energy. In present AA-stacking BLB, only two corner states protected by the $C_{2z}$ symmetry can be observed, differing from the previous reports, e.g., six corner states in $C_{6}$ symmetric graphdiyne~\cite{ref25} and graphyne~\cite{ref27}, three corner states in $C_{3}$-symmetric group-VIB TMDs~\cite{ref28}, and four corner states in $C_{4}$-symmetric $\beta$-$Sb$~\cite{c4symm}.

\begin{figure}[!htb]
    \begin{center}
    \includegraphics[width=\linewidth]{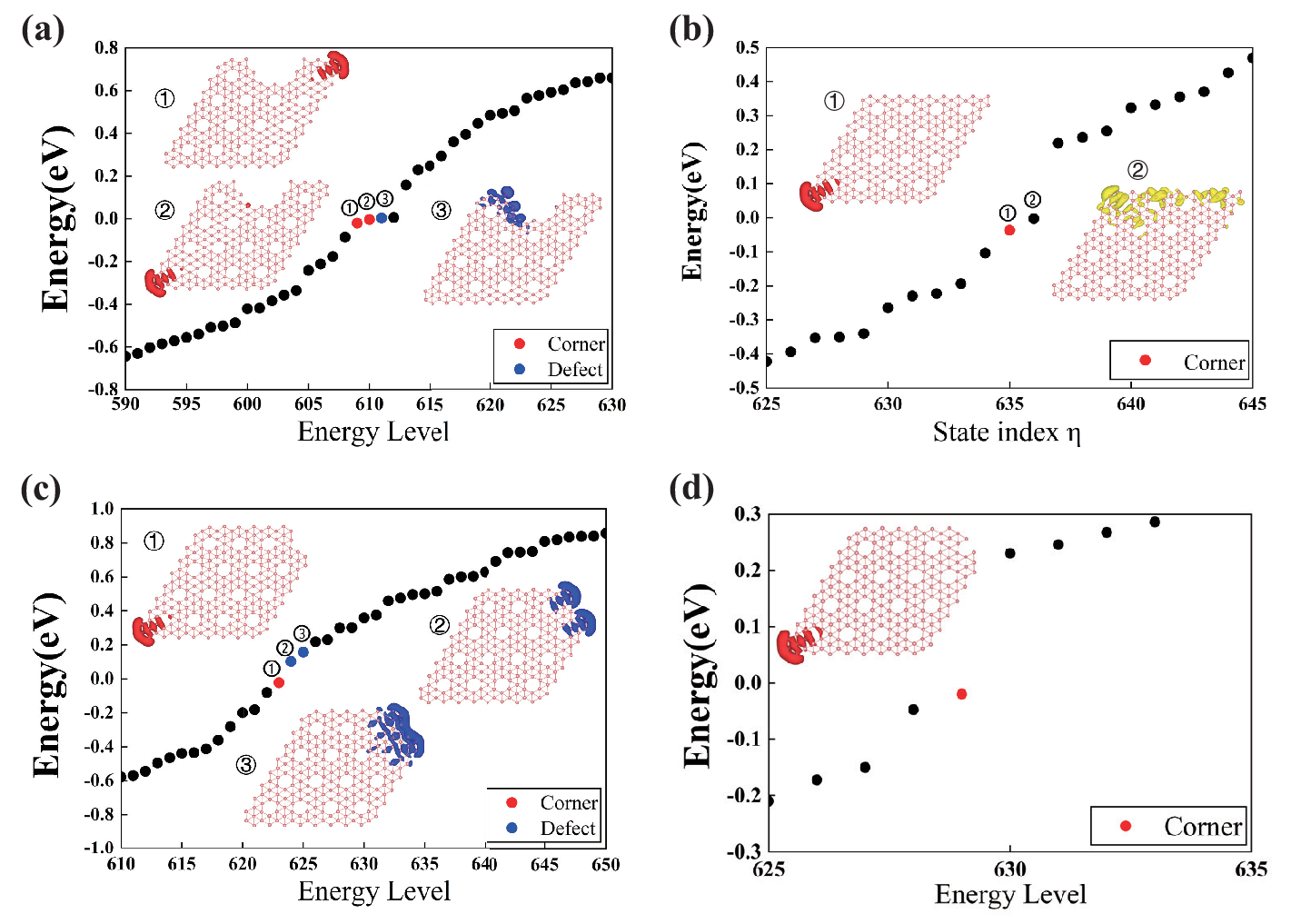}
    \end{center}
	\caption{The energy spectrum of defective AA-stacking $\alpha_{5}$-BLB for nanodisks with ZZ-\textrm{III} edges. (a) A semicircle defect along one selected edge. (b) One additional row removing along one of the edge.  (c) A defect at corner with acute angle. (d) Another type of defects at corner with acute angle. The localized corner states, and defect states are colored by red and blue dots, respectively. }
 \label{Fig6}
\end{figure}

Interestingly, the corner states can appear either at the acute or obtuse angle, similar to the previous reported cases in $C_6$-symmetric porous network model~\cite{ref40}. In principle, the $1D$ edge mode, inherit from the $2D$ bulk, is robust against the boundary configurations even with weak symmetry-breaking perturbation in the first-order $2D$ topological insulators. In contrast, the $0$D end state or junction state, inherited from the $1D$ bulk, is dictated by the terminating unit cell in $1D$ topological insulator~\cite{1d1,1d2,1d3,1d4}. The sensitiveness of $0D$ corner states to the geometries of nanodisks manifests that they inherit from the $1D$ topology of the edge states, further supporting the second-order topology in $\alpha_{5}$-BLB.
\begin{figure}[!htb]
	\begin{center}
		\includegraphics[width=\linewidth]{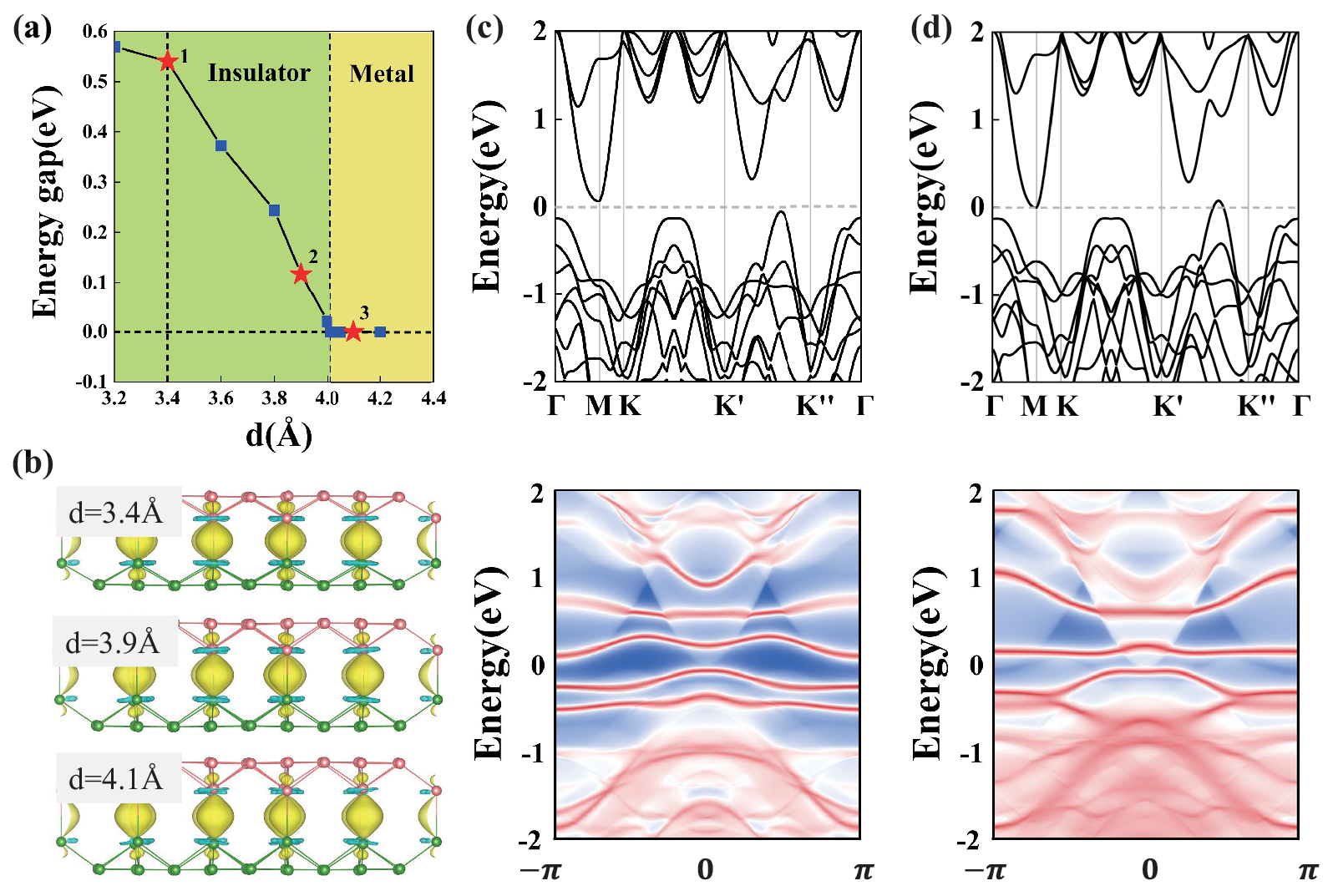}
	\end{center}
	\caption{(a) Band gap as a function of layer spacing $d$ of AA-stacking $\alpha_{5}$-BLB, here $d$ is defined as the distance between two monolayers. The red stars denoted by $1, 2, 3$ correspond to the optimal case ($d=3.4$ \AA), slighltly below, and above the critical value ($d=3.9$ \AA \ and $4.1$ \AA), respectively. (b) The charge differences relative to monolayer case for three $d$ corresponding to the red star $1, 2, 3$, respectively.(c) and (d) are the band structure of the state marked by red star $2$ and $3$, respectively. Upper panels are for the bulk band structure, and lower panels are for the band structure in cylinder geometry. }
	\label{Fig7}
\end{figure}

Furthermore, the influence of different structural defects to the corner states is explored. As shown in Fig.~\ref{Fig6}(a), a semicircle-like defect is introduced on the zigzag edge of the nanodisk but without destroying two acute angles. It is found that the two corner states remain robust through the degeneracy lifting slightly. Besides, a new localized state develops around the defect owing to the appearance of a new acute angle along the ZZ-\textrm{III} edge. Further removing a row of boron atoms leads to the destruction of one initial acute angle of the nanodisk, as a result, the corresponding corner state vanishes (Fig.~\ref{Fig6}(b)). Similar situation can also be observed when the acute angle in the nanodisk is destroyed by introducing two defects, the initial corner state at this acute angle is replaced by two localized defective states at two new generated acute angles (Fig.~\ref{Fig6}(c)). For comparison, another type of defects are introduced at the acute corner but with no emergence of the new defective acute angle, no localized corner or defect states could be found. For completeness, we also calculate the edge states, the energy spectrums of nanodisks and the charge real-space distribution of corner states in the AB- and AC-stacking $\alpha_{5}$-BLBs (see Fig. S5 and Fig. S6). The AB-stacking cases share the similar properties with the AA-stacking one, while the corner states in AC-stacking $\alpha_{5}$-BLB are submerged into the edge states. These results suggest that the corner states are robust against the defects even with $C_{2}$ symmetry breaking. The sensitiveness of corner states, as well as the localized defect states, further supports that these $0$D topological junction/end states stem from the $1D$ topologically nontrivial edge mode.

The sharp difference between the monolayer and bilayer $\alpha_{5}$-borophens provides an avenue to manipulate the topology of borophens. We, then, explored the relation of the electronic properties of these BLBs with the interlayer distance $d$, defined as the distance between two ``vdW'' monolayers. The indirect band gap narrows quickly with the increasing $d$, and vanishes at a critical $d_{c}\sim 4.0$ \AA \ as summarized in the $d$ dependent phase diagram (Fig.~\ref{Fig7}(a)). Specifically, we show two typical cases with (2) $d=3.9$ \AA \ (insulating state just below $d_{c}$) and (3) $d=4.1$ \AA \ (metallic state just above $d_{c}$). Both the bulk band structure and the edge states in case (2) shares the similar features with that in the optimal case (1) ($ d=3.4$ \AA), though the indirect gap much reduced down to $0.11$ eV. Since neither the bulk bands nor the edge state have phase transitions, it still belongs to the same SOTI state, and two degenerate states are found to sit below the Fermi level. In case (3), the indirect gap vanishes leading to a metallic state but with the finite direct gap. The edge states remain robust but submerge into the bulk bands.  Importantly, We notice that the narrowing of indirect band gap is accompanied with the weakened B-B interlayer covalent bond, further manifesting the key role of interlayer B-B bond in gap opening. Such features provide a feasible way to control the electronic structures of bilayer borophene, as well as their topology.

\section{Conclusion}
In summary, based on the first-principle calculations and tight-binding model simulations, we show that the three existing stacked $\alpha_{5}$-BLBs, protected by the $C_{2}$ rotational symmetry, are the $2D$ SOTIs. In contrast to the metallic and thermodynamic unstable monolayer borophene, the formation of interlayer B-B covalent bonds help to stabilize the bilayer counterparts and open the large bulk gap near the Fermi level. The nontrivial second-order topology in AA-stacking BLB is characterized by the  quantized quadrupole momentum $Q_{ij}=\frac{1}{2}$. The SOTI nature is further identified by the gapped and isolated edge states in $1$D ribbon geometry and the emergent corner states  in $0$D nanodisks. These corner states are robust against the structural defects and weak $C_{2}$-symmetry breaking perturbations. The sensitive of corner states to the edge configurations manifests that the second-order topology originates from the $1D$ topology of edge states. Our studies propose the SOTI candidates in realistic materials, and provide the avenue to study the topological properties in borophenes.

{ {\it Acknowledgments}}---This work is supported by National Natural Science Foundation of China (Grant Nos. 12204404 \& 12374137), the Six Talent Peaks Project in Jiangsu Province (XCL-104), the Natural Science Foundation of Jiangsu Province (Grant. No. BK20231397). We thank the computational resources at Yangzhou University.

\bibliography{reference}
\bigskip

\newpage
\section*{Appendix}
\setcounter{figure}{0}
\setcounter{equation}{0}
\renewcommand \thefigure{S\arabic{figure}}
\renewcommand \theequation{S\arabic{equation}}

\begin{figure*}[!htb]
	\begin{center}
		\includegraphics[width=0.7\linewidth]{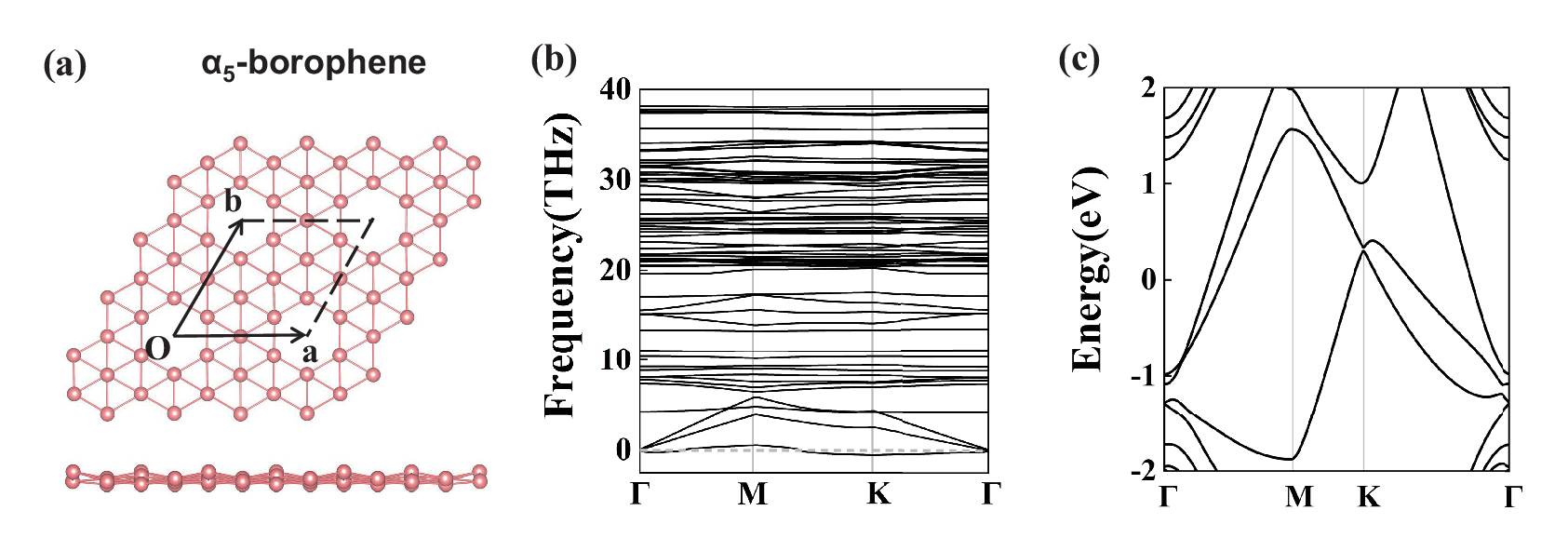}
	\end{center}
	\caption{(a) The optimized lattice structure of $\alpha_{5}$-phase boron monolayer. (b) The phonon spectrum with imaginary frequencies around $K$ point. (c) Band structure of monolayer $\alpha_{5}$-phase boron monolayer along high symmetry paths in Brillouin zone. }
	\label{Fig.S1}
\end{figure*}

\begin{figure*}[!htb]
	\begin{center}
		\includegraphics[width=0.7\linewidth]{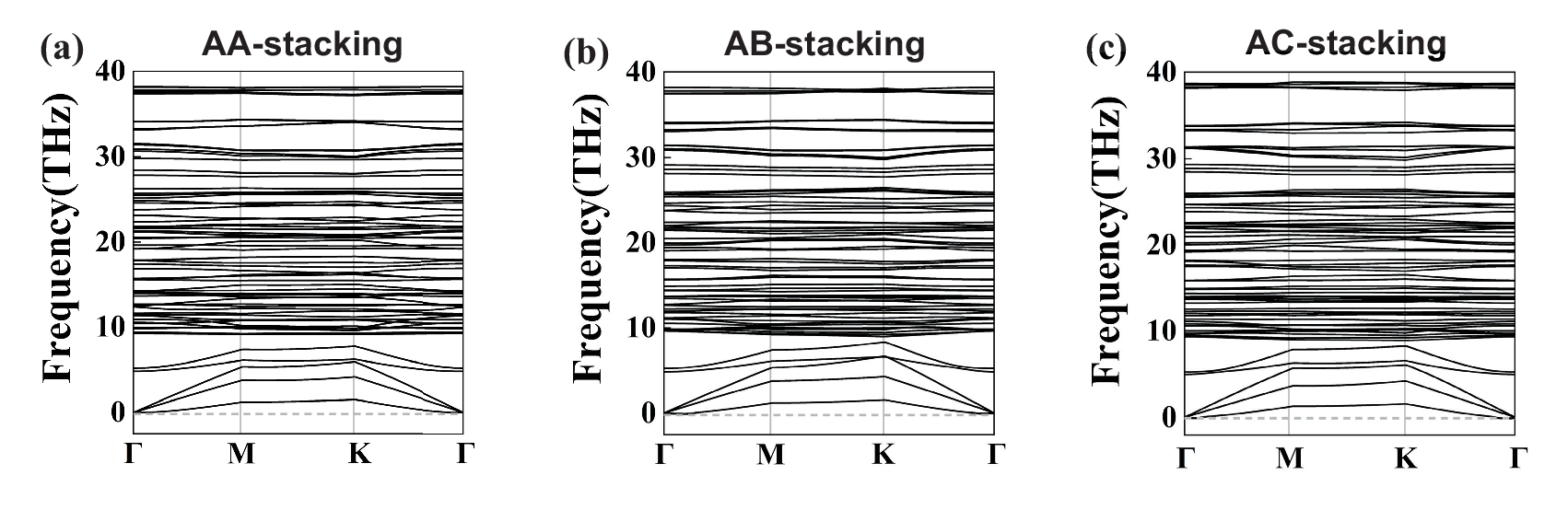}
	\end{center}
	\caption{The phonon spectrums of (b) AA, (c) AB and (d) AC stacking $\alpha_{5}$-BLB along high symmetry paths in Brillouin zone. No imaginary frequencies are observed. }
	\label{Fig.S2}
\end{figure*}

\begin{figure*}[!htb]
	\begin{center}
		\includegraphics[width=0.7\linewidth]{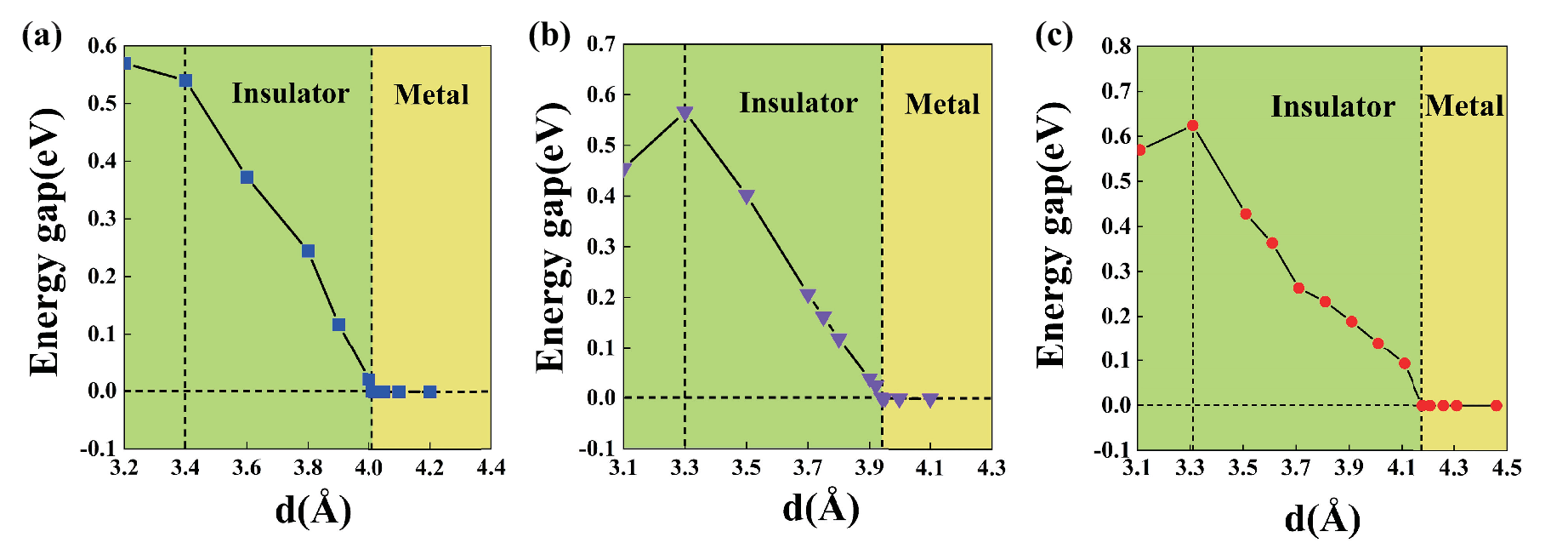}
	\end{center}
	\caption{The evolution of band gaps and increase of layer spacing $d$, (a) AA, (b) AB and (c) AC stacking $\alpha_{5}$-BLB, respectively. The green and yellow areas indicate insulator and metal phase. }
	\label{Fig.S3}
\end{figure*}

\begin{table*}[!htb]
\caption{\label{tab:table4}%
	Parities of 33 occupied bands at $\Gamma$ and three M points for AA stacking $\alpha_{5}$-BLB. The label + and - denotes even and odd parity at TRIM, respectively. }
\begin{ruledtabular}
\begin{tabular}{p{3cm}<{\centering}p{15cm}<{\centering}}
	TRIM&Parity\\
	\hline
	$\Gamma$&$+ + + - + + - - + - - - + + + - - + - - + - - - + - - + + - + - +$\\
	$M_{1}$&$+ - + - + + - - + - + - + - + - + - + - - - + + - - + - + + + - -$\\
	$M_{2}$&$+ - + - + + - - + - + - + - + - + - + - - - + + - - + - + + + - -$\\
	$M_{3}$&$- - + + - - + + + + + - - + - - - + + + + - + - - + - - + - + + +$
\end{tabular}
\end{ruledtabular}
\end{table*}

\begin{figure*}[!htb]
	\begin{center}
		\includegraphics[width=0.7\linewidth]{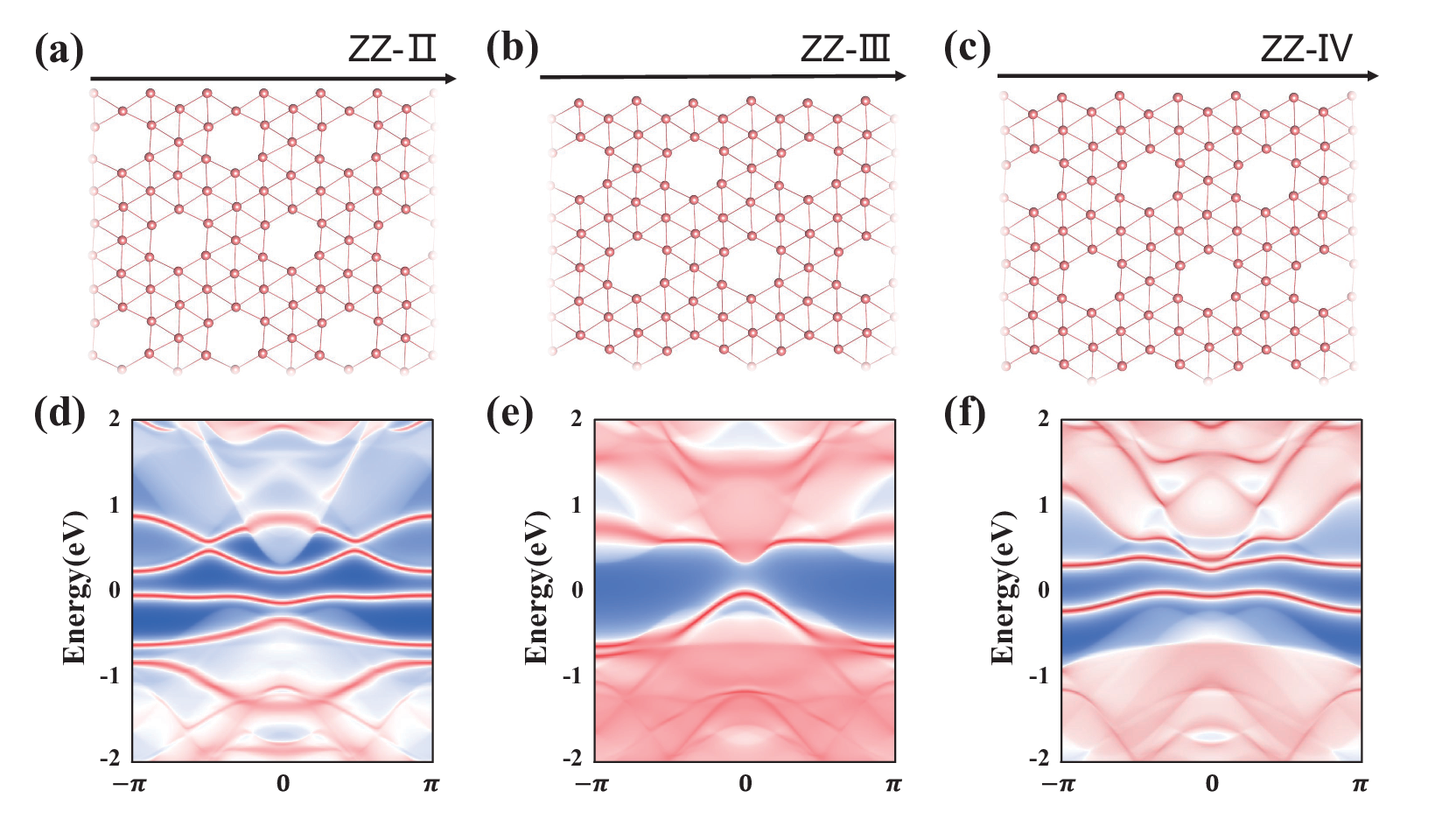}
	\end{center}
	\caption{(a)-(c) Top view of the structure of AA-stacking $\alpha_{5}$-BLB. The black lines mark three types of zigzag edges. (d)-(f) The edge states of semi-infinite planes with ZZ-\text{\textrm{II}}, ZZ-\text{\textrm{III}} and ZZ-\text{\textrm{IV}} edge, calculated by the Wannier interpolation method.}
	\label{Fig.S5}
\end{figure*}

\begin{figure*}[!htb]
	\begin{center}
		\includegraphics[width=0.7\linewidth]{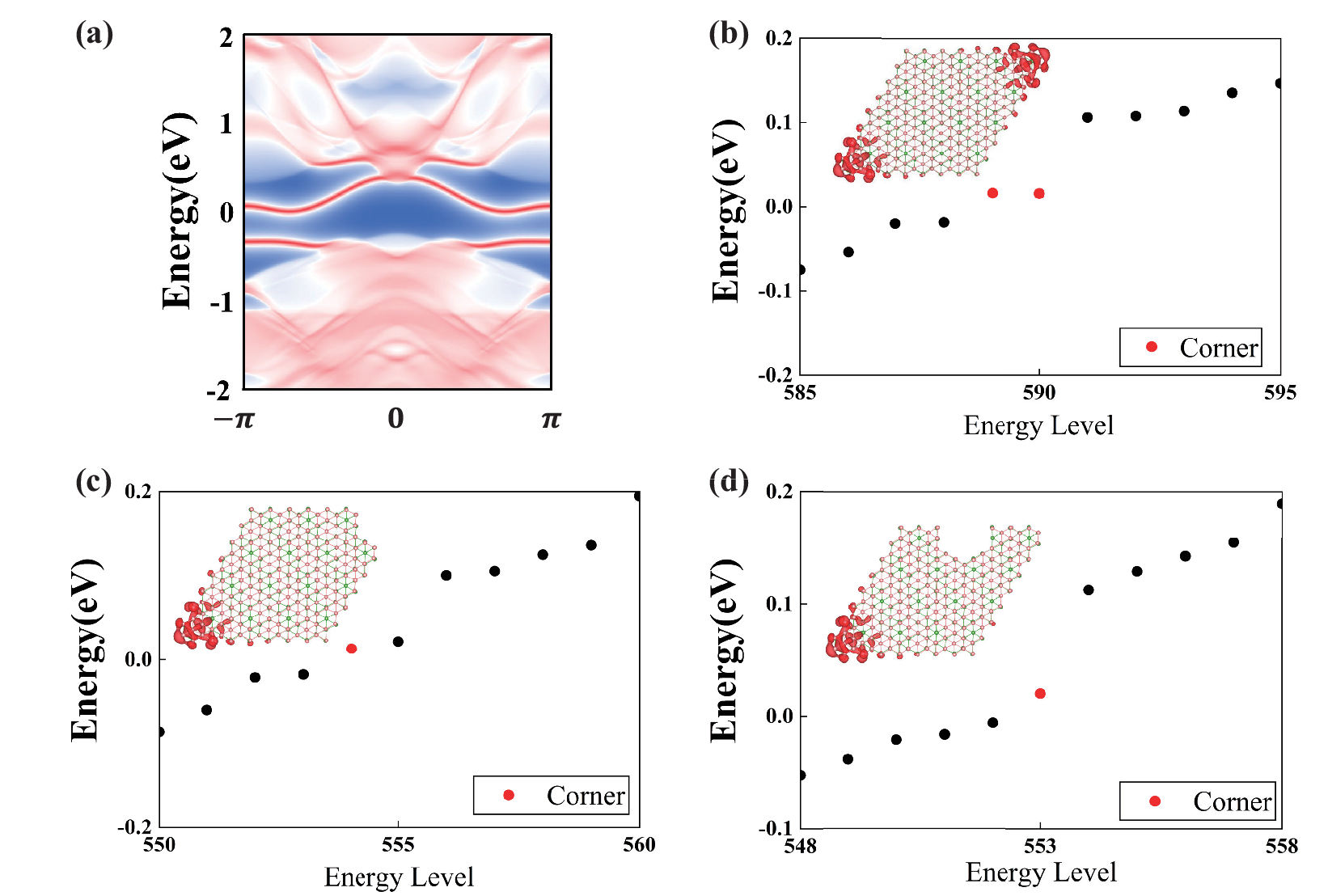}
	\end{center}
	\caption{(a) The edge states of AB- stacking $\alpha_{5}$-BLB calculated by the Wannier interpolation method. The energy spectrum of three nanodisks calculated by VASP, the corner states are marked by red dots. The insets are the real-space distribution of the corner states. (b) Nanodisk without defects. (c) Nanodisk with a defect at one acute part. (d) Nanodisk with a defect at one edge.}
\label{Fig.S7}
\end{figure*}

\begin{figure*}[!htb]
	\begin{center}
		\includegraphics[width=0.7\linewidth]{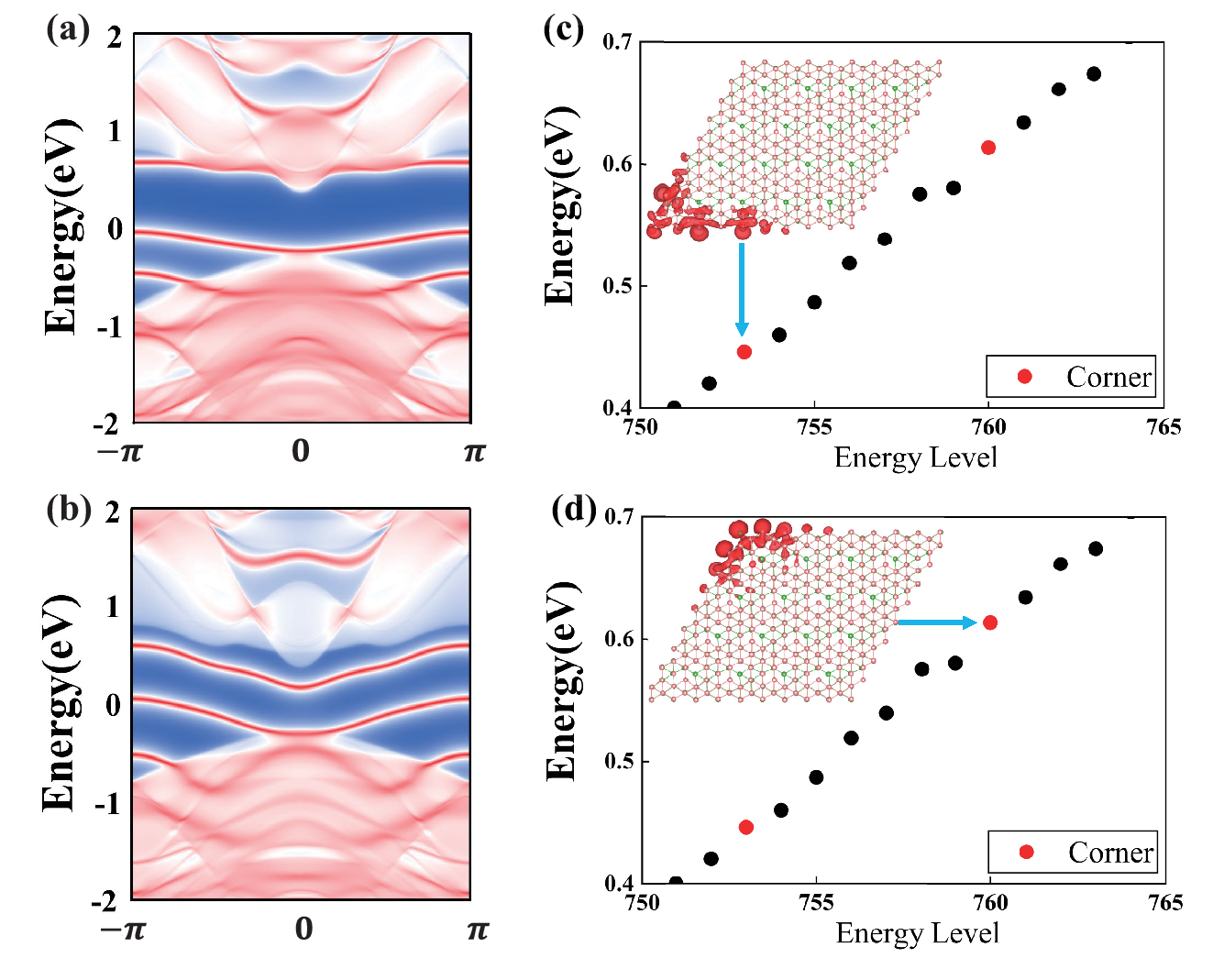}
	\end{center}
	\caption{(a) and (b) Two types of edge states of AC- stacking $\alpha_{5}$-BLB calculated by the Wannier interpolation method. (c) and (d) The energy spectrum of a nanodisk calculated by VASP, the corner states are marked by red dots. The insets are the real-space distribution of the corner states.}
	\label{Fig.S8}
\end{figure*}

\begin{figure*}[!htb]
	\begin{center}
		\includegraphics[width=0.7\linewidth]{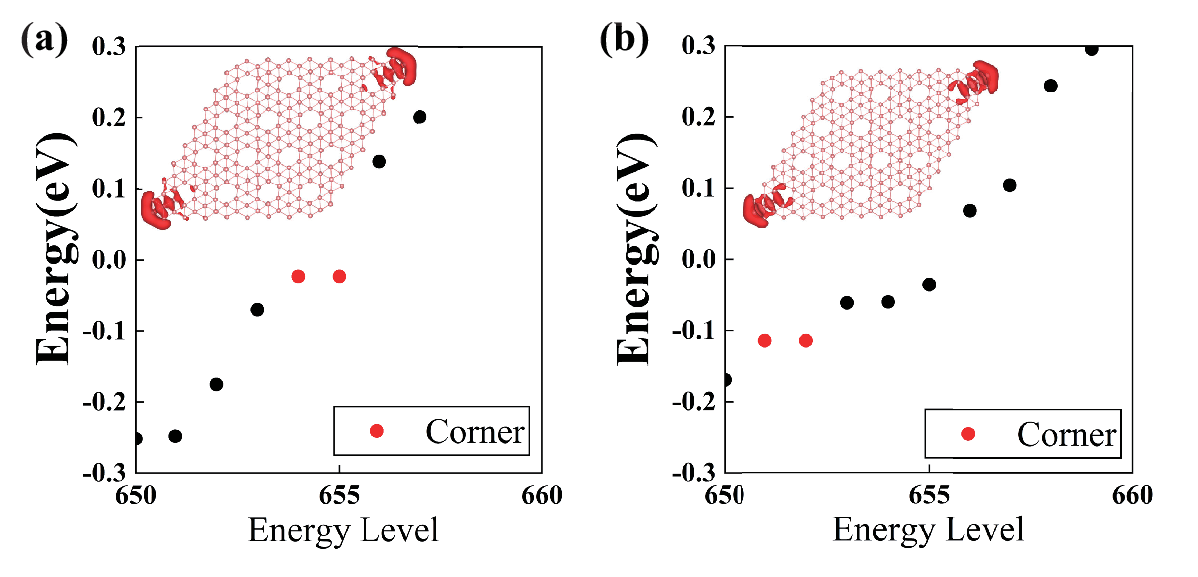}
	\end{center}
	\caption{The energy spectrum of the nanodisk of (a) $d=3.9$ \AA \ and (b) $d=4.1$ \AA, the corner states are marked by red dots. The inset is the real-space distribution of the corner states.}
	\label{Fig.S9}
\end{figure*}

\end{document}